\documentclass[%
 reprint,
 amsmath,amssymb,
 aps,
]{revtex4-2}

\usepackage{graphicx}
\usepackage{dcolumn}
\usepackage{bm}
\usepackage{siunitx}
\usepackage{color}
\usepackage{autobreak}
\usepackage{braket}
\usepackage{here}
\usepackage{array}
\usepackage{appendix}
\usepackage[normalem]{ulem}

\begin{document}

\preprint{APS/123-QED}

\title{Elastic study of electric quadrupolar correlation in the paramagnetic state of a frustrated quantum magnet  $\rm Tb_{2+\it \delta}Ti_{2-\it \delta}O_7$} 

\author{Y. Nii$^{1,2}$}
\author{Y. Hirokane$^{3}$}
\author{S. Nakamura$^{1}$}
\author{S. Kimura$^{1}$}
\author{Y. Tomioka$^{4}$}
\author{T. Nojima$^{1}$}
\author{Y. Onose$^{1}$}

\affiliation{
$^{1}$Institute for Materials Research, Tohoku University, Sendai 980-8577, Japan.\\
$^{2}$PRESTO, Japan Science and Technology Agency (JST), Kawaguchi 332-0012, Japan.\\
$^{3}$Department of Basic Science, University of Tokyo 153-8902, Japan.\\
$^4$National Institute of Advanced Industrial Science and Technology (AIST), Tsukuba 305-8562, Japan
}


\begin{abstract}
Electric quadrupolar state in a frustrated quantum magnet $\rm Tb_{2+\it \delta}Ti_{2-\it \delta}O_7$ has been studied by means of ultrasonic and magnetostriction measurements. A single crystal showed elastic anomaly at about 0.4 K, manifesting a long-range quadrupole ordering. By investigating anisotropy of the magnetoelastic responses, we found a crossover temperature for the strongly correlated quadrupole state, below which the experimental data of elastic constant and magnetostriction become qualitatively different from their calculations based on a single-ion model. We suppose that relatively high onset temperature of the quadrupole correlation compared with the transition temperature is ascribed to the geometrical frustration effect, and this correlated state seems to be responsible for the unusual properties in the paramagentic state of $\rm Tb_2Ti_2O_7$.
\end{abstract}

\maketitle

\section{Introduction}
Geometrically frustrated systems have been serving as a fertile playground for studying nontrivial magnetic phenomena \cite{anderson1987resonating, balents2010spin}. The pyrochlore lattice is a prototypical structure having geometrical frustration, where conventional magnetic order fails to develop and non-trivial states often emerge. In particular, $\rm Dy_2Ti_2O_7$ and $\rm Ho_2Ti_2O_7$ exhibit a novel phenomena of spin ice, in which magnetic moments remain disordered down to lowest temperature, showing a zero-point entropy \cite{ramirez1999zero}. Dynamics of magnetic monopoles under magnetic fields have also attracted much attention  \cite{castelnovo2008magnetic}. Distinct from these classical Ising spin systems, the presence of quantum fluctuations, or transverse interaction, in systems with weaker magnetic anisotropy provides more exotic physics related to quantum spin liquid (QSL) state in the ground state \cite{lee2012generic, onoda2011quantum, savary2012coulombic}. In quests of such quantum spin ice systems, intensive studies have been performed \cite{gardner2010magnetic, gingras2014quantum, rau2019frustrated} based on compounds of $\rm Yb_2Ti_2O_7$, $\rm Er_2Ti_2O_7$, $\rm Tb_2Ti_2O_7$, $\rm Pr_2Zr_2O_7$, and so on.\par
Among them, $\rm Tb_2Ti_2O_7$ shows unique magnetic properties with $\rm Tb^{3+}$ ions having a total angular momentum $J$ = 6. The trigonal crystalline electric field (CEF) partially lifts the degeneracy, and produces a low-energy level scheme, that consists of the ground state doublet and first excited doublet separated only by $\approx$ 18 K \cite{gingras2000thermodynamic}. The small gap allows admixing between them, and quantum fluctuation transverse to their local Ising axis becomes important \cite{molavian2007dynamically, lee2012generic, onoda2011quantum}. This material has been thought as a candidate of QSL. Despite a negative Curie-Weiss temperature of $\Theta_{CW}$ = $- 19$ K \cite{gardner2010magnetic,gardner1999cooperative}, no magnetic long-range order was observed down to the achievable lowest temperatures \cite{gardner1999cooperative, gardner2003dynamic}, while short-range magnetic correlation develops below several tens of Kelvin as revealed by various methods: $\mu$SR \cite{dunsiger2003comparison, gingras2000thermodynamic}, AC susceptibility \cite{gardner2003dynamic, hamaguchi2004low}, neutron spin echo\cite{gardner2003dynamic, gardner2004spin}, and diffuse neutron scattering\cite{rule2006field} suggested the characteristic magnetic state often referred to as a cooperative (or correlated) paramagnet \cite{gardner1999cooperative}. In addition to a magnetic dipole moment, these doublets carry a large electric quadrupolar moment, which couples to local strain. Recently, quadrupole order has been discerned around 0.5 K \cite{Takatsu_2016, elastic_2020} for Tb-rich $\rm Tb_{2+\it \delta}Ti_{2-\it \delta}O_7$ crystals. These findings  implies  that quadrupolar  moment  plays  an  important  role in this system. Since quadrupole moment $O_\Gamma$ and strain $e_\Gamma$ are linearly coupled by a quadrupole-strain coupling constant $g_\Gamma$ ($\Gamma$ represents an irreducible representation) as \cite{luthi2007physical}
\begin{align}
\mathcal{H}_{QS} = g_{\Gamma} O_\Gamma e_{\Gamma},\label{Hqs11}
\end{align}
$\rm Tb_2Ti_2O_7$ shows prominent magnetoelastic responses such as significant elastic softening \cite{mamsurova1988anomalies, nakanishi2011elastic, elastic_2020}, giant magnetostriction \cite{mamsurova1988anomalies, pukhov1985crystal, ruff2010magnetoelastics}, suppressed thermal conductivity \cite{li2013phonon}, thermal Hall effect \cite{hirschberger2015thermal, hirokane2019phononic}, and dynamical hybridization of phonon and crystalline electric field (CEF) states \cite{Fennell_2014, ruminy2016sample, constable2017double}. These experiments suggest that pure spin models are not enough to capture the physics of $\rm Tb_2Ti_2O_7$, and quadrupolar degree of freedom and its coupling to the lattice has to be considered carefully.\par
Here we have investigated the quadrupole correlation in $\rm Tb$-rich  $\rm Tb_{2+\it \delta}Ti_{2-\it \delta}O_7$ ($\delta$ $\approx$ 0.009) by measuring the detailed temperature and magnetic field dependences of  elastic constant and magnetostriction. The elastic constant \cite{luthi2007physical} and the strain \cite{klekovkina2011simulations} are expressed as\begin{align}
 C_{\Gamma} &=C_\Gamma^0 -N g_{\Gamma}\left.
\displaystyle\frac{\partial \braket{O_{\Gamma}}}{\partial e_{\Gamma}} \right| _{e_{\Gamma}=0},  \\
    e_{\Gamma} &= \displaystyle\frac{Ng_{\Gamma}}{C_{\Gamma}}\braket{O_{\Gamma}}.
\end{align}Here $C_\Gamma^0$ is the elastic constant without considering quadrupole-strain coupling, $\braket{O_\Gamma}$ represents statistical average of quadrupolar moment, and $N$ is number of $\rm Tb$ ions in a unit volume. These equations mean that two methods in this study are complementary and suitable for probing quadrupole fluctuation and quadrupole moment, respectively. By performing ultrasound measurement down to below 0.4 K, we have found a clear anomaly in elastic constant at the quadruople ordering temperature of 443 mK, being consistent with a previous works  \cite{Takatsu_2016,elastic_2020}. We show strong quadrupolar correlation persists up to $\approx$ 10 K by comparing experimental data and calculation.
\section{Experiment}
A single crystal of $\rm Tb_2Ti_2O_7$ was grown by the floating-zone method under 1 atm $\rm O_2$ atmosphere.
Laue photographs were used to determine the crystallographic orientation. X-ray diffractometer (XRD) measurements (Smart Lab, RIGAKU) was also performed to determine the lattice parameter. 
We also confirmed single-crystallinity and the absence of impurity phase in our sample by these measurements. The sample was cut and carefully polished to obtain rectangular shape with flat and parallel $\{100\}$ surfaces. \par
The size of the sample for ultrasonic measurements is $4.2 \times 3.3 \times 1.7$ mm$^3$. Polyvinylidene difluoride (PVDF) films with thickness of 9 $\rm\mu m$ was attached on [001] surfaces by room-temperature-vulcanizing (RTV) silicon. Longitudinal ultrasound of about 100 MHz is generated/detected by these PVDF transducers. Based on the pulse-echo method, elastic constant of $C_{11}$, where both wavevector $\textbf{k}$ and polarization vector $\textbf{u}$ are parallel to [001], was obtained. Ultrasound echoes and reference signal were directly converted to digital data using an oscilloscope (RTO1004, ROHDE $\&$ SCHWARZ), and phase comparison analysis was numerically performed. Note that magnetostriction of $\rm Tb_2Ti_2O_7$ was so large that adhesion between transducer and sample was easily broken under large magnetic fields when a hard transducer such as LiNbO$_3$ was used. Therefore, we used the organic PVDF film that can be applicable even under large magnetostriction. Only the longitudinal mode is available for the PVDF transducer.\par
Magnetostriction measurement was carried out by a conventional strain-gauge technique.
We simultaneously measured longitudinal and transverse strains by orthogonally attaching two strain-gauges with external magnetic field applied along [001]. In order to eliminate false strain due to magnetoresistance of strain-gauge, reference sample (glass) was also measured by the same settings. 
True magnetostriction was derived by subtracting the magnetoresistance from raw data. Ultrasonic measurement below 2 K was performed by using a $^3$He refrigerator at Center for Low Temperature Science in Tohoku University. The lowest temperature achieved was 0.38 K. Detailed magnetic field angle dependence was measured above 2 K using a $^4$He refrigerator equipped with a rotation stage.

\begin{figure}
    \centering
    \includegraphics[width=3.375in]{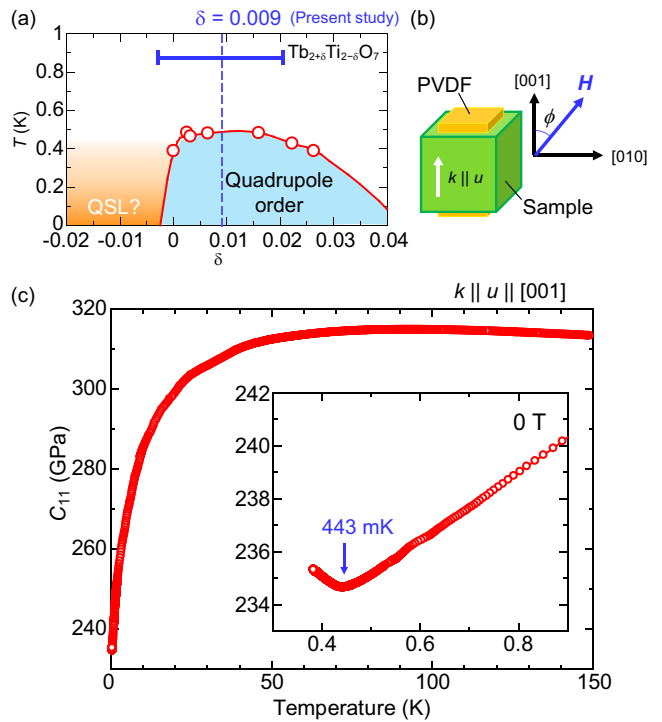}
    \caption{(a) $\delta-T$ phase diagram of $\rm Tb_{2+\it \delta}Ti_{2-\it \delta}O_7$ . The composition of present study is estimated to $\delta \approx 0.009$. Horizontal error bar indicates the uncertainty of $\delta$.  Circles were data taken from Ref. \cite{Wakita_2016}. (b) Experimental set-up of ultrasound measurement. (c) Temperature dependence of compressive elastic constant $C_{11}$ at 0 T. Deflection of $C_{11}$ at 443 mK suggests the ordering of quadrupole moment.
    }
    \label{fig1}
\end{figure}
\section{Results}
As shown in Fig. 1(a), stoichiometric $\rm Tb_{2}Ti_{2}O_7$  locates close to the boundary of a QSL and a quadrupolar ordered phase, and the ground state is very sensitive to the off-stoichiometric parameter $\delta$ \cite{Taniguchi_2015,Wakita_2016}. Our crystal was found to be slightly Tb rich having $\delta \approx 0.009$ (see Appendix for the determination of $\delta$). Therefore, the ground state may show a quadrupolar ordering. Figure 1 (c) exhibits temperature dependence of compressive elastic constant $C_{11}$. Large elastic softening was observed with decreasing temperature followed by the hardening below 443 mK. The softening is caused by quadrupole fluctuation while the hardening suggests that the ground state degeneracy was lifted by a phase transition. This up-turn is consistent with the previous reports \cite{elastic_2020}, and can be attributable to the quadrupolar ordering transition \cite{Takatsu_2016}. \par
\begin{figure*}
    \centering
    \includegraphics[width=6.75in]{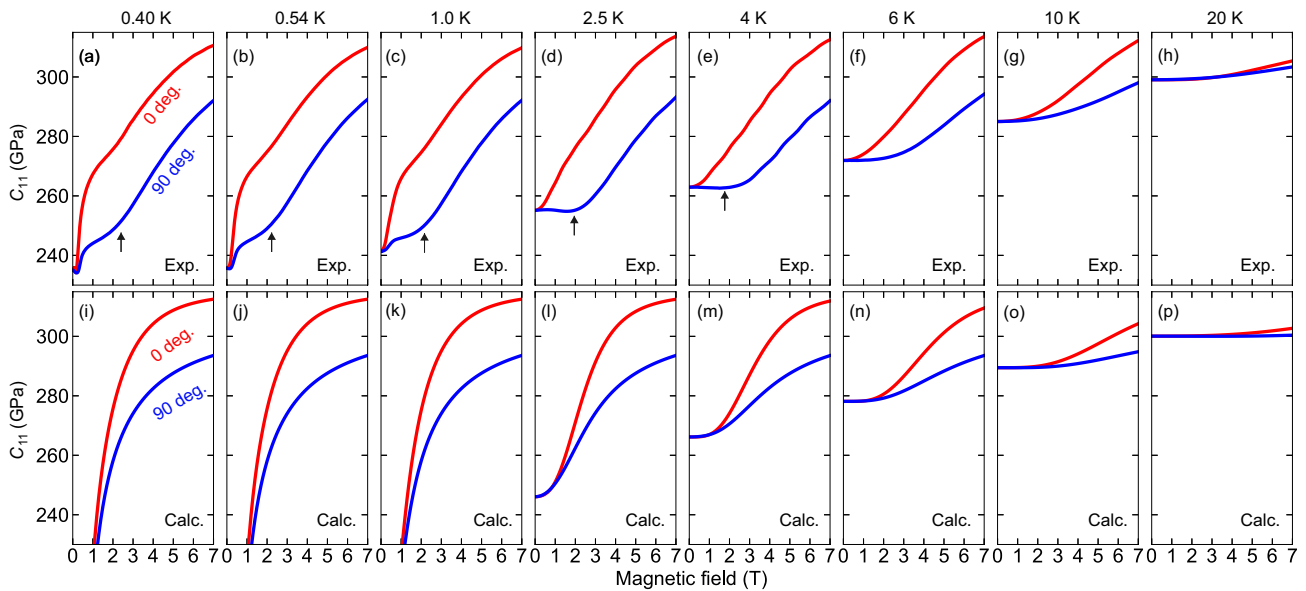}
    \caption{Magnetic field dependence of (a)-(h) experimental and (i)-(p) calculated elastic constant $C_{11}(\phi)$ at various temperatures under magnetic field applied parallel ($\phi$= 0 deg.) and perpendicular ($\phi$= 90 deg.) to the propagation direction of ultrasound.}
    \label{fig2}
\end{figure*}
In order to investigate how the quadrupole correlation evolves toward the transition temperature, we have studied the elastic constant $C_{11}$ in the para-quadrupole state. Figures 2 (a)-(h) show $C_{11}$ at various temperatures under magnetic field applied parallel ($\phi = 0$ deg.) or perpendicular ($\phi = 90$ deg.) to the propagation direction of ultrasound [see Fig. 1 (b) for the definition of $\phi$]. When the magnetic field is applied, the quadrupole fluctuation is suppressed and $C_{11}$ increases owing to the polarization of magnetic dipole moment. The polarization also induces the anisotropy of elastic constant. The magnitude of anisotropy is larger in the higher magnetic field and lower temperature. In particular, $C_{11}$ shows large anisotropy below 4 K; $C_{11}(\phi = 0 \ \rm deg.)$ steeply increases with increasing magnetic field, but $C_{11}(\phi = 90 \ \rm deg.)$ remains nearly constant below 2 T.  Below 1 K, a small cusp was observed for $\phi = 90$ deg. in the low magnetic field region. To show the effect of quadrupole correlation, we have performed the calculation of elastic constant based on a single-ion model as shown in Figs. 2(i)-2(p). In this calculation,  CEF, Zeeman field, and the quadrupole-strain coupling were taken into account, while any interaction between $\rm Tb^{3+}$ moments was not (the details of the calculation procedure are shown in the Appendix B). The difference between the experimental data and the calculation should be ascribed to the correlation effect. The calculated elastic modulus at 0 T diverges toward the zero-temperature following the Curie-type elastic softening \cite{luthi2007physical}. As a result, the elastic constant shows unphysical negative value below 0.45 K. This is partly because higher-order elastic constants and inter-quadrupolar coupling are neglected. Appart from the negative divergence, the calculated $C_{11}$ of the two orthogonal configurations almost isotropic below 2 T irrespective temperature. On the other hand, the experimental elastic constant shows large anisotropy in the low temperature region while the experiment and the calculation are similar above 10 K.   In particular, $C_{11}$  below 4 K shows gradual kinks as indicated by arrows only for the 90 deg. experimental data. The signatures different from the single-ion calculation seem to be owing to the quadrupole correlation effect.\par
In order to investigate the elastic anisotropy at low magnetic fields in more detail, we have systematically studied magnetic field angle dependence of  $\Delta C_{11}(\phi) = C_{11}(\phi) - C_{11}(\phi = 0\ \rm deg.)$. 
We show experimentally measured and calculated angular dependences under various magnetic fields at a fixed temperature of 4 K in Figs. 3(a) and 3(b), respectively.
In a low field, the experimental data of elastic constant vary with the angle as  $\cos 2 \phi$ while the calculation does as $\cos 4 \phi$. The $\phi = 0$ and $90$ deg. data are different for the $\cos 2 \phi$ dependence while they are same for the $\cos 4 \phi$ dependence. Therefore, the difference between the experimental and calculation is consistent with that shown in Fig. 2. As the magnetic field is increased, both the angular dependences become similar to each other.\par
Figures 4(a) and 4(b) show experimental and calculated $\Delta C_{11}(\phi)$ at 1 T for various temperatures. Clear angle dependences gradually appear upon cooling for both the experimental and calculated data. On the other hand, the main part of experimentally observed angular dependence contains $\cos 2 \phi$ dependent component while the calculation shows $\cos 4 \phi$ dependence, which is consistent with the results in Fig. 3 at low field. Note that this discrepancy does not merely originates from the unoptimized parameters used in our calculation. The $\cos 2 \phi$ dependence cannot be reproduced even when the quadrupolar-strain coupling parameters ($g_{22}$ and $g_{yz}$) were varied over a wide range. Thus, the $\cos2\phi$  dependence can be viewed as a measure of quadrupole correlation, and it is useful to estimate the onset temperature of correlation. To evaluate these angular dependences, we fitted them by $\Delta C_{11}(\phi) = A_{2\phi} \cos 2\phi + A_{4\phi} \cos 4\phi$, where $A_{2\phi}$ and $A_{4\phi}$ are fitting parameters at 1 T. Figures 5(a) and 5(b) represent temperature dependence of these fitting parameters. The most apparent disagreement between experiment and calculation appears in $A_{2\phi}$ below $\approx$ 10 K.
This onset temperature of quadrupole correlation is quite large compared with the quadrupolar-ordering transition temperature of $\approx$ 0.4 K.\par
\begin{figure}
    \centering
    \includegraphics[width=3.1 in]{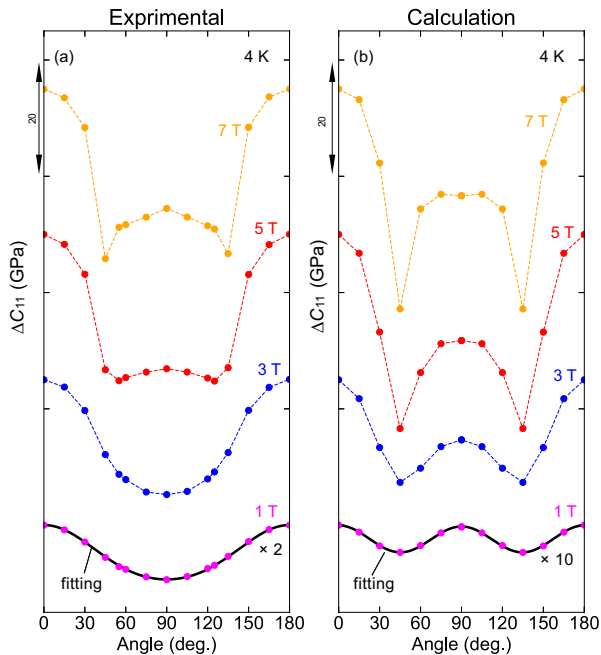}
    \caption{Magnetic field angle dependence of (a) experimental and (b) calculated $\Delta C_{11}(\phi)$ under various magnetic fields at 4 K. Magnetic field is rotated with respect to the [100] axis as shown in Fig. 1 (b). Data are shown with offset for clarity. Experimental and calculated data at 1 T are multiplied by 2 and 10, respectively. The black solid lines represent fitting curves of $\Delta C_{11}(\phi) = A_{2\phi} \cos 2\phi + A_{4\phi} \cos 4\phi$ at 1 T .}
    \label{fig3}
\end{figure}
\begin{figure}
    \centering
    \includegraphics[width=3.1 in]{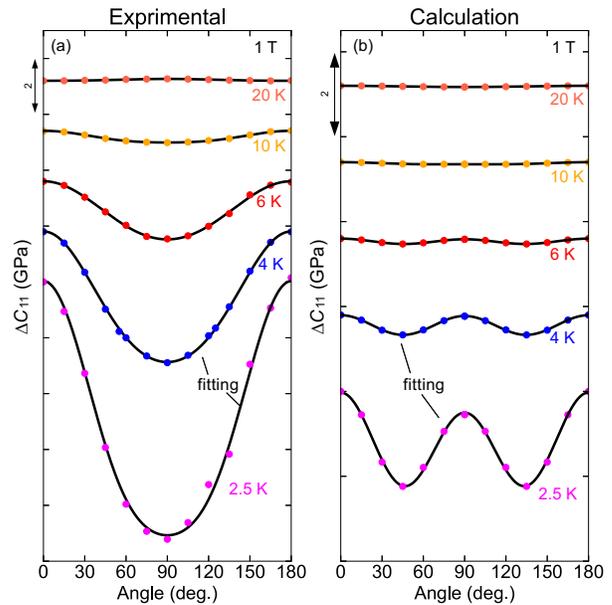}
    \caption{Magnetic field angle dependence of (a) experimental and (b) calculated $\Delta C_{11}(\phi)$ at various temperatures. Magnetic field is as large as 1 T and rotated with respect to the [100] axis as shown in Fig. 1 (b). Data are shown with offset for clarity. Each solid line represents fitting curves of $\Delta C_{11}(\phi) = A_{2\phi} \cos 2\phi + A_{4\phi} \cos 4\phi$.}
    \label{fig3}
\end{figure}
\begin{figure}
    \centering
    \includegraphics[width=3.1 in]{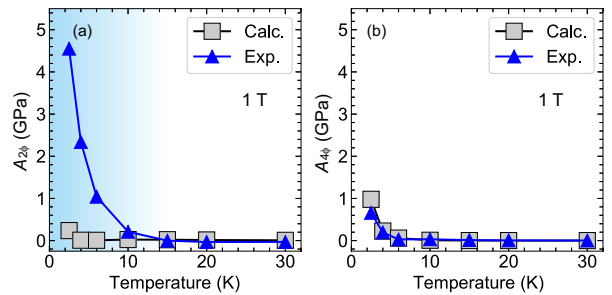}
    \caption{Temperature dependence of fitting parameters (a) $A_{2\phi}$ and (b) $A_{4\phi}$ for the experimental (blue triangle) and the calculation (gray square) at 1 T.}
    \label{fig3}
\end{figure}

To investigate the correlation effect from a different probe, we have measured magnetic field induced strain, i.e., magnetostriction. Complementary to elastic response, magnetostriction probes the quadrupolar moment induced by magnetic field. 
Figure 6(a) shows magnetostriction measured at various temperatures. 
Magnetic field was applied along the [001] direction, and the longitudinal ($e_{zz}\ ||$ [001]) and the transverse ($e_{xx}\ ||$ [100]) strains were measured simultaneously [see Fig. 6(b)]. 
Being consistent with the previous studies \cite{pukhov1985crystal, mamsurova1988anomalies, ruff2010magnetoelastics}, $e_{zz}$ is positive while $e_{xx}$ is negative, and the magnitudes were as large as order of $10^{-3}$. As the temperature is decreased, the magnitude of magnetostriction gradually increases. This is reflected by the evolution of magnetic susceptibility.
We have also calculated the magnetostrictions for both configurations using the same single-ion model and parameters as in the case of elastic constant (see Appendix B). 
It well reproduces the experimental data in the high temperature region, but discrepancy between the experiment and calculation becomes apparent in the low temperature region below 10 K. This can also be ascribed to the quadrupole correlation effect. To evaluate this discrepancy, the difference between experiment and calculation was exhibited in Fig. 6(c). Similar to the previous case of elastic constant, it evolves rapidly below around 10 K. The most obvious discrepancy is the magnetostoriction perpendicular to the magnetic field; the magnitude of experimental data is almost half of the calculated value. In this direction, the experimental elastic constant is also small compared with the calculation. These discprepancies correspond to each other and are considered to be related to quadrupolar correlation.\par
\begin{figure}
    \centering
    \includegraphics[width=3in]{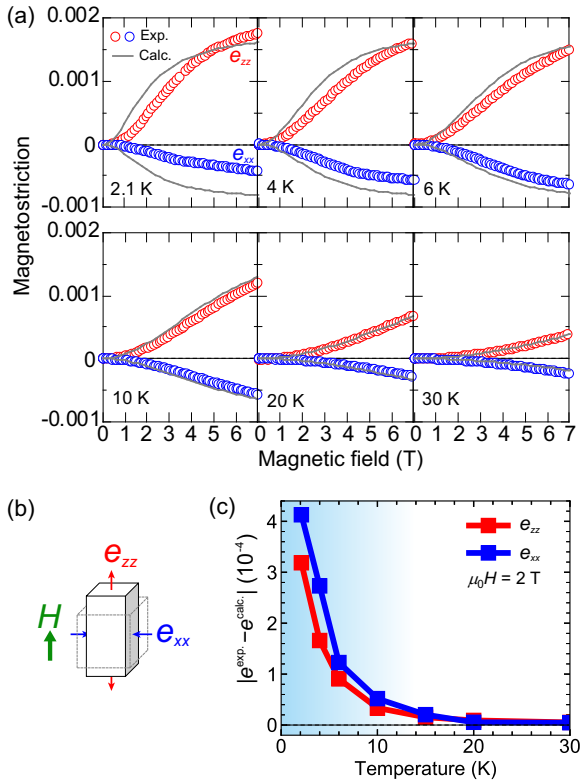}
    \caption{(a) Magnetostriction at various temperatures. $e_{zz}$ (red) or $e_{xx}$ (blue) represent strains parallel or perpendicular to the magnetic field, respectively. The external magnetic field is applied along [001]. The gray lines represent calculated magnetostriction for both configurations. (b) Experimental configuration of magnetostriction measurements. (c) Difference of magnetostriction between experiment and calculation at a fixed magnetic field of 2 T.}
    \label{fig4}
\end{figure}
\begin{figure}
    \centering
    \includegraphics[width=3.1in]{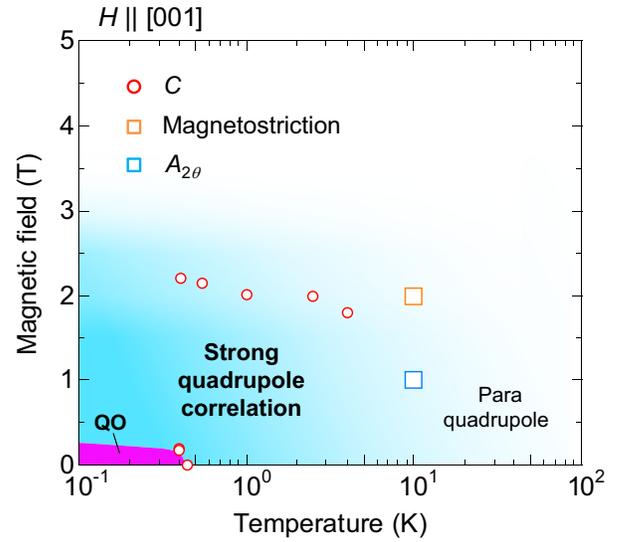}
    \caption{Phase diagram of $\rm Tb_{2+\it \delta}Ti_{2-\it \delta}O_7$ ($\delta$ = 0.009). QO stands for a long-range quadrupolar ordered phase. Squares indicate the onset temperature of the quadrupolar correlation determined by $A_{2\phi}$ and magnetostriction. Red circles represent deflection points of $C_{11}$ shown by arrows in Fig. 2(a)-(e). The boundary fields between the strong correlation and induced magnetic state should be larger than them.}
    \label{fig5}
\end{figure}
\section{Discussion and Conclusion}
In conclusion, we have revealed strong quadrupole correlation in a frustrated quantum magnet $\rm Tb_{2+\it \delta}Ti_{2-\it \delta}O_7$ by means of ultrasound and magnetostriction measurements. Figure 7 summarizes the obtained results. Consistent with the previous reports \cite{Takatsu_2016,elastic_2020}, our Tb-rich single crystal showed elastic anomaly at about 0.4 K, manifesting a long-range quadrupole ordering. We have compared the observed elastic constant and magnetostriction with the calculated data based on a single-ion model.
The discrepancy between the observed and calculated data was discerned below about 10 K, indicating breakdown of single-ion picture and strong quadrupole correlation in the paramagnetic state. The onset temperature is far above the long-range quadrupolar  ordering temperature of 443 mK.  The magnetic field range of the strong correlation is not clear but supposedly Tesla order judging from the magnetic field dependences of elastic constant [Figs. 2(a)-2(e), 3(a) and 3(b)] and magnetostriction [Fig. 6(a)].
In the correlated region, the difference between experiment and calculation was confirmed clearly in the response along the direction perpendicular to the magnetic field. In this direction, the magnetostriction (i.e., $e_{xx}$) becomes negative and the elastic response (at $\phi$ = 90 deg.) remains soft at low magnetic field. These imply that quadrupolar moment tends to avoid the perpendicular direction and quadrupolar fluctuation persists under magnetic field. Although previous theoretical studies took the quadrupole degree of freedom and its correlation into accounts \cite{onoda2011quantum,Takatsu_2016}, the related magnetoelastic responses were not reported so far. The observations in this study seem useful for the future development of theoretical investigation.\par
Previous experimental studies also suggested the importance of quadrupolar correlation. Ruff $et\ al.$ observed a substantial broadening of Bragg peaks below 20 K, and attributed it to structural fluctuation caused by the quadrupolar correlation \cite{ruff2007structural}.  Inelastic neutron scattering experiments showed that the first excited CEF mode becomes dispersive below about 20 K, suggesting CEF levels among adjacent $\rm Tb^{3+}$ ions are interacted \cite{guitteny2013anisotropic, Fennell_2014}. Moreover,  in a sister compound  $\rm Tb_2Ge_2O_7$, recent comprehensive study showed that significant correlations between quadrupolar moments of $\rm Tb^{3+}$ ions are present above 1.1 K \cite{hallas2020intertwined}. All these suggests that the onset of quadrupolar correlation is much higher than the long-range quadrupolar ordering temperature.\par
The anomalous physical properties in the paramagnetic state such as giant thermall Hall effect \cite{hirschberger2015thermal,hirokane2019phononic}, magnetoelastic hybrid excitation \cite{Fennell_2014,ruminy2016sample}, and vibronic state between phonon and CEF states \cite{constable2017double} may be related to the effect of quadrupole correlation. While the strong quadrupolar correlation effect persists up to several tens of Kelvin, the geometrical frustration seems to suppress the long-range quadrupolar ordering similarly to  other pyrochlore systems. The characteristic of $\rm Tb_{2}Ti_{2}O_7$ clarified in this work is that the strong correlation effect shows up in the sector of quadrupole or lattice dynamics.

\section*{Acknowledgement}
We are grateful to S. Onoda for fruitful discussions. This work is supported by JSPS KAKENHI (grant numbers JP20K03828, JP21H01036), and PRESTO (grant number JPMJPR19L6), and the Mitsubishi Foundation.
\appendix
\renewcommand{\thefigure}{A\arabic{figure}}
\setcounter{figure}{0}
\renewcommand{\thetable}{A\arabic{table}}
\setcounter{table}{0}
\section{Off-stoichiometric parameter x}
As reported in Ref. \cite{Taniguchi_2015, Wakita_2016}, the off-stoichiometric parameter $\delta$ of $\rm Tb_{2+\it \delta}Ti_{2-\it \delta}O_7$ is proportional to lattice parameter $a$. Thus, precise determination of the lattice parameter allows us to estimate $x$. Here we performed XRD measurement using a powder sample crushed from a piece of a single crystal. Followed by the calibration of standard Si powder sample and least-square fitting of the Bragg peaks, we determined the lattice parameter of the sample to 10.1539 $\pm$ 0.0015 $\rm \AA$. Then, $\delta$ = 0.009 $\pm$ 0.012 was deduced according to a relation: $a(\rm \AA) = 0.124418\ \it x \rm + 10.15280$ \cite{Wakita_2016}.

\begin{figure}
    \centering
    \includegraphics[width=3.375in]{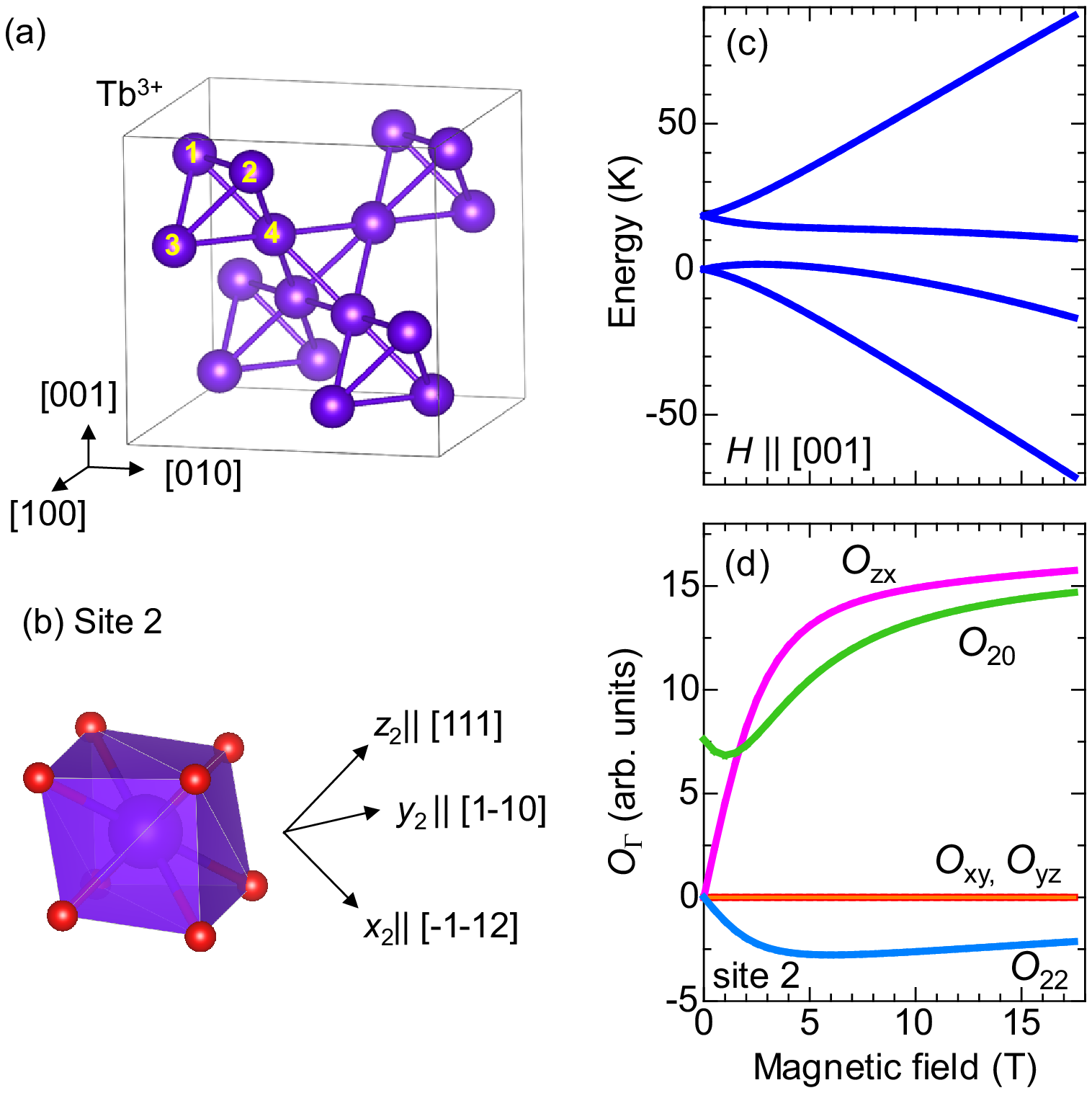}
    \caption{(a) A schematic of $\rm Tb^{3+}$ ions in $\rm Tb_2Ti_2O_7$ forming pyrochlore lattice. (b) Definition of local crystal axes in a $\rm TbO_8$ polyhedron at site 2 in (a). (c) Magnetic field dependence of low-energy CEF levels. The level schemes of every four sites are equivalent when magnetic field is applied along [001]. (d) Magnetic field dependence of electric quadrupole moments at the site 2.}
    \label{figA1}
\end{figure}
\section{Single-ion based calculation}
Here we show the details of calculation for elastic constant and magnetostriction. It is based on a single-ion model without intersite interaction. As shown in Fig. A1(a), $\rm Tb^{3+}$ ions have four inequivalent sites on a pyrochlore lattice. There is no interaction between them in this model. To obtain the elastic constant and magnetostriction, we separately calculated these quantities for each site and then averaged the four values.
\subsection{Local coordinate}
The local coordinates ($\bm{x}_m,\ \bm{y}_m,\ \bm{z}_m$) of the four sites $m = $1, 2, 3, 4 are defined using the global axes as shown in Table ~\ref{tab:table1} and Figs. A1(a) and A1(b).
\begin{table}[b]
\caption{\label{tab:table1}
Correspondence between local orthonormal axes and global cubic axes. $m$ stands for the four inequivalent $\rm Tb^{3+}$ sites shown in Fig. A1(a)
}
\begin{ruledtabular}
\begin{tabular}{cccc}
 $m$&
$\bm{x}_m$&
$\bm{y}_m$&
$\bm{z}_m$\\
\colrule
        1 & $\frac{1}{\sqrt{6}}(1,1,2)$ & $\frac{1}{\sqrt{2}}(-1,1,0)$ & $\frac{1}{\sqrt{3}}(-1,-1,1)$\\
        2 & $\frac{1}{\sqrt{6}}(-1,-1,2)$ & $\frac{1}{\sqrt{2}}(1,-1,0)$ & $\frac{1}{\sqrt{3}}(1,1,1)$\\
        3 & $\frac{1}{\sqrt{6}}(-1,1,-2)$ & $\frac{1}{\sqrt{2}}(1,1,0)$ & $\frac{1}{\sqrt{3}}(1,-1,-1)$\\
        4 & $\frac{1}{\sqrt{6}}(1,-1,-2)$ & $\frac{1}{\sqrt{2}}(-1,-1,0)$ & $\frac{1}{\sqrt{3}}(-1,1,-1)$\\
\end{tabular}
\end{ruledtabular}
\end{table}
Using this relation, physical tensors represented by local axes at $m$-site can be transformed. For instance, the strain tensors $e_{ij}^m$ described at a local coordinate $m$ and the strain tensor $e_{ij}$ defined in the global frame are transformed by $e_{\alpha \beta}^m = \sum_{i,j} R_{\alpha i}^m R_{\beta j}^m e_{ij}$, where the rotation matrices $R^m$ are given by
\begin{align}
R^1 &= 
\left(
\begin{array}{ccc}
 \frac{1}{\sqrt{6}} & -\frac{1}{\sqrt{2}} & -\frac{1}{\sqrt{3}} \\
 \frac{1}{\sqrt{6}} & \frac{1}{\sqrt{2}} & -\frac{1}{\sqrt{3}} \\
 \sqrt{\frac{2}{3}} & 0 & \frac{1}{\sqrt{3}} \\
\end{array}
\right),\nonumber\\
R^2 &= 
\left(
\begin{array}{ccc}
 -\frac{1}{\sqrt{6}} & \frac{1}{\sqrt{2}} & \frac{1}{\sqrt{3}} \\
 -\frac{1}{\sqrt{6}} & -\frac{1}{\sqrt{2}} & \frac{1}{\sqrt{3}} \\
 \sqrt{\frac{2}{3}} & 0 & \frac{1}{\sqrt{3}} \\
\end{array}
\right),\nonumber\\
R^3 &= 
\left(
\begin{array}{ccc}
 -\frac{1}{\sqrt{6}} & \frac{1}{\sqrt{2}} & \frac{1}{\sqrt{3}} \\
 \frac{1}{\sqrt{6}} & \frac{1}{\sqrt{2}} & -\frac{1}{\sqrt{3}} \\
 -\sqrt{\frac{2}{3}} & 0 & -\frac{1}{\sqrt{3}} \\
\end{array}
\right),\nonumber\\
R^4 &= 
\left(
\begin{array}{ccc}
 \frac{1}{\sqrt{6}} & -\frac{1}{\sqrt{2}} & -\frac{1}{\sqrt{3}} \\
 -\frac{1}{\sqrt{6}} & -\frac{1}{\sqrt{2}} & \frac{1}{\sqrt{3}} \\
 -\sqrt{\frac{2}{3}} & 0 & -\frac{1}{\sqrt{3}} \\
\end{array}
\right).\nonumber
\end{align}

\subsection{Hamiltonian}
The local Hamiltonian at $m$-site ($m$ = 1-4) are given by
\begin{equation}
    \mathcal{H}^m = \mathcal{H}_{CEF}^m + \mathcal{H}_Z^m + \mathcal{H}_{QS}^m,\label{Hamiltonian}
\end{equation}
where
\begin{align}
    &\mathcal{H}_{CEF}^m = B_{20}O_{20}^m+B_{40}O_{40}^m+B_{43}O_{43}^m+B_{60}O_{60}^m\label{Hcef}\\
    &\quad+B_{63}O_{63}^m+B_{66}O_{66}^m,\nonumber\\
    &\mathcal{H}_{Z}^m= -g_J \mu_ B \mu_0 \bm{J}^m\cdot\bm{H},\label{Hz}\nonumber\\
    &\mathcal{H}_{QS}^m = \sum_{\Gamma} g_{\Gamma} O_{\Gamma}^m e_{\Gamma}^m.\nonumber
\end{align}
Here $\mathcal{H}_{CEF}^m$, $\mathcal{H}_{Z}^m$, $\mathcal{H}_{QS}^m$ are Hamiltonians representing CEF, Zeeman energy, quadrupolar-strain coupling, respectively. $B_{ij}$ and $O_{ij}^m$ are CEF parameters and the Stevens operators at $m$-site. We have taken CEF parameters from Ref. \cite{ruminy2016crystal} (see Table~\ref{tab:table2}).
\begin{table}[b]
\caption{\label{tab:table2}
The $B_{ij}$ parameters used in the present study. The parameters were taken from Ref. \cite{ruminy2016crystal} and shown in meV units. In Ref. \cite{ruminy2016crystal}, CEF Hamiltonian is defined as $\mathcal{H}_{CEF} = \sum_{k,q} \tilde{B}^k_q C^k_q$, where $C^k_q$ is Wybourne tensor operators. $\tilde{B}_q^k$ is converted to $B_{i,j}$ by using relations of $B_{20} = \alpha_J \lambda_2^0 \tilde{B}^2_0$, $B_{4q} = \beta_J \lambda_2^q \tilde{B}^4_q$, and $B_{6q} = \gamma_J \lambda_2^q \tilde{B}^6_q$, where $\alpha_J = -1/99$, $\beta_J = 2/16335$, $\gamma_J = -1/891891$, $\lambda_2^0 = 1/2$, $\lambda_4^0 = 1/8$,
$\lambda_4^3 = \sqrt{35}/2$, $\lambda_6^0 = 1/16$, $\lambda_6^3 = \sqrt{105}/8$, and $\lambda_6^6 = \sqrt{231}/16$.
}
\begin{ruledtabular}
\begin{tabular}{ccc}
        $B_{20}$ & $B_{40}$ & $B_{43}$ \\
\colrule
        -0.282323 & 0.00474441 & 0.0412876 \\
    \hline \hline
        $B_{60}$ &$B_{63}$ &$B_{66}$ \\
    \colrule
        -4.51288$\times 10^{-6}$ & 0.000120922 & -0.000137393\\
\end{tabular}
\end{ruledtabular}
\end{table}
$g_J$ = 3/2, $\mu_B$, $\mu_0$, $\bm J \it^m$, and $\bm H$ represent Lande’s g-factor, Bohr magneton, permeability in a vacuum, total angular momentum and magnetic field, respectively.
$O_{\Gamma}^m$, $e_{\Gamma}^m$, and $g_{\Gamma}$ are irreducible quadrupolar operator, strain tensor, and the coupling constant belonging to the symmetry $\Gamma$, respectively. Here $\bm J \it^m$, $O_{ij}^m$ , $O_{\Gamma}^m$, and $e_{\Gamma}^m$ are defined in its local frame at $m$-site. The explicit form of the Stevens operators is given by \cite{hutchings1964point}
\begin{align}
	&O_{20}^m = 3(J_z^m)^2-J(J+1),\nonumber\\
	&O_{40}^m = 35 (J_z^m)^4 - 30 J(J+1)(J_z^m)^2 + 25(J_z^m)^2 \nonumber\\
	&\quad-6J(J+1)+ 3J^2(J+1)^2,\nonumber\\
	&O_{43}^m = \Bigl\{J_z\bigl[(J_+^m)^3 + (J_-^m)^3\bigr] \nonumber\\
	&\quad +\bigl[(J_+^m)^3 + (J_-^m)^3\bigr]J_z\Bigr\}/4,\nonumber\\
	&O_{60}^{(m)} = 231 (J_z^m)^6-315J(J+1)(J_z^m)^4 + 735(J_z^m)^4\nonumber\\
	&\quad + 105 J^2 (J+1)^2 (J_z^m)^2 - 525 J(J+1)(J_z^m)^2\nonumber\\
	&\quad+ 294(J_z^m)^2-5 J^3 (J + 1)^3 + 40 J^2 (J + 1)^2\nonumber\\
	&\quad - 60 J (J + 1),\nonumber\\
	&O_{63}^m =\Bigl\{\bigl[11 (J_z^m)^3 - 3 J (J + 1) J_z^m - 59 J_z^m\bigr]\nonumber \\
	&\quad \times \bigl[(J_+^m)^3 + (J_-^m)^3\bigr]+ \bigl[(J_+^m)^3 + (J_-^m)^3\bigr]\nonumber\\
	&\quad \times \bigl[11 (J_z^m)^3 - 3 J (J + 1) J_z^m - 59 J_z^m\bigr] \Bigr\}/4,\nonumber\\
    &O_{66}^m = \bigl[(J_+^m)^6 + (J_-^m)^6\bigr]/2.\nonumber
\end{align}
Considering the local $D_{3d}$ symmetry at the $\rm Tb^{3+}$ site, $\mathcal{H}_{QS}$  can be reduced as
\begin{align}
\mathcal{H}_{QS}^m &= \sum_{\Gamma} g_{\Gamma} O_{\Gamma}^m e_{\Gamma}^m\nonumber\\
&= g_{20}O_{20}^me_{20}^m+ g_{22}\bigl(O_{22}^me_{22}^m + O_{xy}^me_{xy}^m\bigr)\nonumber \\
&\quad+ g_{yz}\bigl(O_{yz}^me_{yz}^m + O_{zx}^me_{zx}^m\bigr),\nonumber
\end{align}
where irreducible representations of $O_{\Gamma}^m$, and $e_{\Gamma}^m$ are given by
\begin{subequations}
\begin{align}
&O_{20}^m = 3(J_z^m)^2-J(J+1),\nonumber\\
&O_{22}^m = \big[(J_x^m)^2-(J_y^m)^2\big]/\sqrt{2},\nonumber\\
&O_{xy}^m = (J_x^mJ_y^m+J_y^mJ_x^m)/\sqrt{2},\nonumber\\
&O_{yz}^m = (J_y^mJ_z^m+J_z^mJ_y^m)/\sqrt{2},\nonumber\\
&O_{zx}^m = (J_z^mJ_x^m+J_x^mJ_z^m)/\sqrt{2},\nonumber\\
&e_{B}^m = e_{xx}^m + e_{yy}^m + e_{zz}^m,\nonumber\\
&e_{20}^m= (2e_{zz}^m-e_{xx}^m-e_{yy}^m)/3,\nonumber\\
&e_{22}^m = (e_{xx}^m-e_{yy}^m)/\sqrt{2}.\nonumber
\end{align}
\end{subequations}
By diagonalizing the Hamiltonian shown in Eq.(\ref{Hamiltonian}), eigenstates and eigen energies are obtained. Figure A1(c) shows the obtained eigen energies as a function of magnetic field. Although only the four CEF levels are represented, all the thirteen eigen states are taken into accounts in the following calculation.

\subsection{Elastic constant}
To get the magnetoelastic responses, total free energy of $F = F_{elas} + F_{elec}$ is considered \cite{luthi2007physical}. The $F_{elas} = \sum_{i,j,k,l}(1/2)C^0_{ijkl}e_{ij}e_{kl}$ is the elastic part of the free energy and $F_{elec} =- (N/\beta)\ln Z$ is the electronic part of the free energy, respectively. Here $C_{ijkl}^0$, $N$ and  $\beta = 1/k_BT$ represent elastic stiffness tensors without electronic contribution, total number of Tb$_4$ tetrahedrons in a unit volume, and inverse temperature, respectively. $Z = \sum_{n,m} e^{-\beta E_{n,m}}$  is partition function, where $E_{n,m}$ is $n$-th eigen energy at $m$-site. Then $C_{11}$ is given by taking a second derivative of the total free energy with respect to strain $e_{zz}$ as \cite{luthi2007physical, klekovkina2011simulations}
\begin{align}
C_{11} &= \displaystyle\frac{\partial^2 F}{\partial e_{zz}^2}\nonumber\\ &= C_{11}^0 - N \sum_{m=1}^4\sum_{\Gamma} g_{\Gamma} \displaystyle\frac{\partial \braket{O_{\Gamma}^m}}{\partial e_{zz}}
\cdot\Bigl(\displaystyle\frac{\partial e_{\Gamma}^m}{\partial e_{zz}}\Bigr).
\end{align}

\begin{figure}
    \centering
    \includegraphics[width=3.375in]{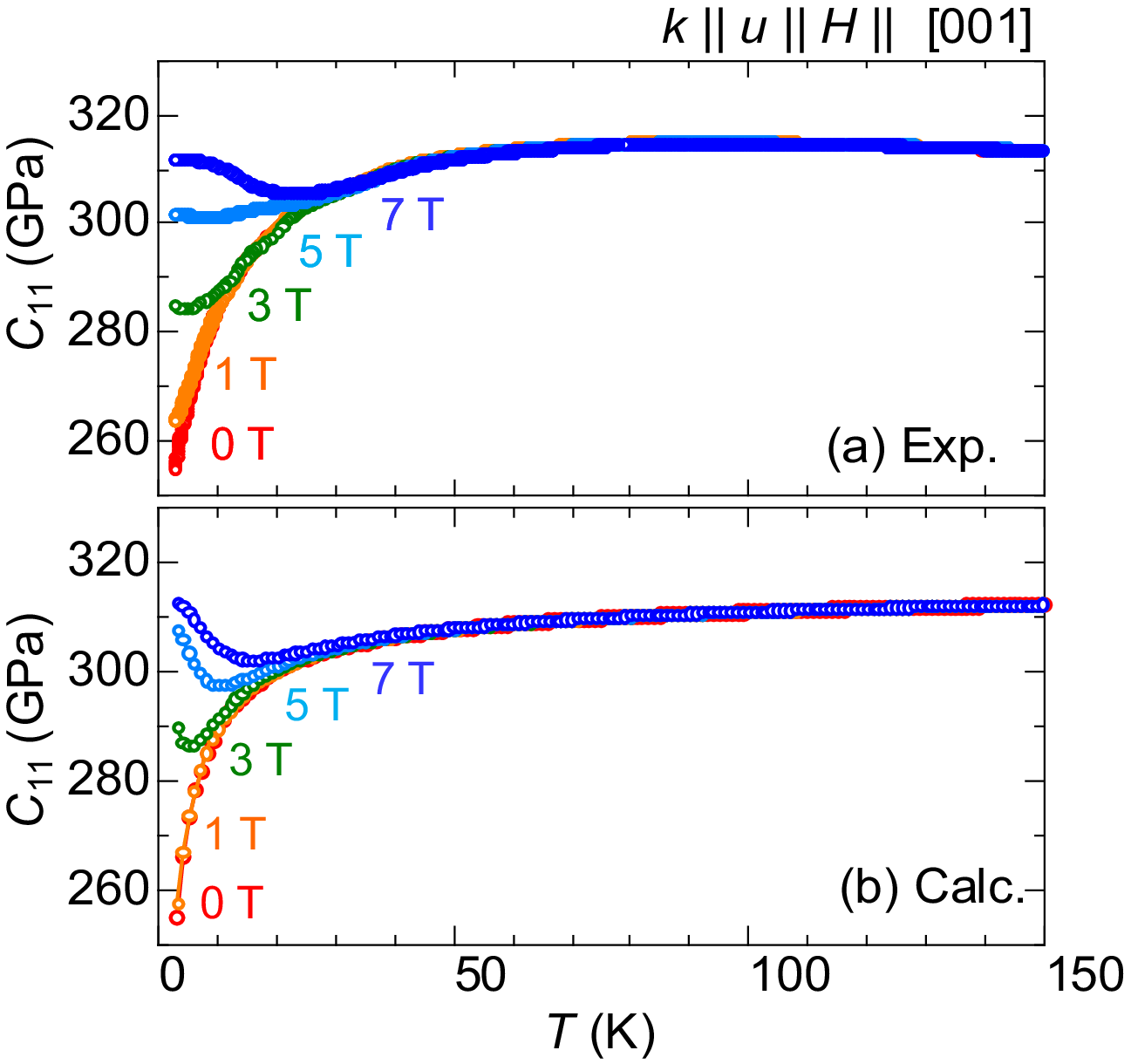}
    \caption{Temperature dependence of (a) experimental and (b) calculated elastic constants $C_{11}$ above 2 K at various magnetic fields along [001].}
    \label{figA2}
\end{figure}

The first term is the background elastic constant corresponding to the contribution from the purely elastic energy, and the second term is the modification by the quadrupole-strain coupling. For simplicity, $C_{11}^0$ is approximated to be constant and temperature independent. The $\braket{O}$ represents the statistical average of the quadrupole moment defined as $\braket{O} = \sum_n \braket{n|O|n}e^{-\beta E_n}$. Here $\ket{n}$ are $n$-th eigenstates. The derivative term $\partial \braket{O_\Gamma^m}/\partial e_{zz}$ was approximated to $(\braket{O_\Gamma^m}_{\Delta e_{zz}}-\braket{O_\Gamma^m}_{0})/\Delta e_{zz}$, where $\Delta e_{zz}$ was set to small finite value of the order of $10^{-6}$. The $\braket{O_\Gamma^m}_{\Delta e_{zz}}$ and $\braket{O_\Gamma^m}_{0}$ means $\braket{O_\Gamma^m}$ with and without the small $\Delta e_{zz}$. Figure A2 shows comparison of experimental and calculated $C_{11}$ as a function of temperature.  Our calculation reproduces the overall behavior including a broad dip at above 3 T. This supports our calculation and estimated parameters are reasonable.

\subsection{Magnetostriction}
By minimizing the free energy with respect to the strain tensors \cite{pukhov1985crystal, klekovkina2011simulations} (i.e., $\partial F/\partial e_{xx} = \partial F/\partial e_{yy} = \partial F/\partial e_{zz} = 0
$), one gets coupled equations as follows:
\begin{align}
&C_{11}^0 e_{xx} + C_{12}^0 (e_{yy} + e_{zz}) = N\sum_{m=1}^4\sum_{\Gamma} g_{\Gamma} \braket{O_{\Gamma}^m}\cdot\Bigl(\displaystyle\frac{\partial e_{\Gamma}^m}{\partial e_{xx}}\Bigr),\nonumber\\
&C_{11}^0 e_{yy} + C_{12}^0 (e_{zz} + e_{xx}) = N\sum_{m=1}^4\sum_{\Gamma} g_{\Gamma} \braket{O_{\Gamma}^m}\cdot\Bigl(\displaystyle\frac{\partial e_{\Gamma}^m}{\partial e_{yy}}\Bigr),\nonumber\\
&C_{11}^0 e_{zz} + C_{12}^0 (e_{xx} + e_{yy}) = N\sum_{m=1}^4\sum_{\Gamma} g_{\Gamma} \braket{O_{\Gamma}^m}\cdot\Bigl(\displaystyle\frac{\partial e_{\Gamma}^m}{\partial e_{zz}}\Bigr).\nonumber
\end{align}
Then, three strain tensors can be given by 
\begin{align}
e_{xx} &= S_0 \Bigl[(C_{11}^0 + C_{12}^0)\tilde{\sigma}_{xx} - C_{12}^0 (\tilde{\sigma}_{yy} + \tilde{\sigma}_{zz})\Bigr],\nonumber\\
e_{yy} &= S_0 \Bigl[(C_{11}^0 + C_{12}^0)\tilde{\sigma}_{yy} - C_{12}^0 (\tilde{\sigma}_{zz} + \tilde{\sigma}_{xx})\Bigr], \nonumber\\
e_{zz} &= S_0 \Bigl[(C_{11}^0 + C_{12}^0)\tilde{\sigma}_{zz} - C_{12}^0 (\tilde{\sigma}_{xx} + \tilde{\sigma}_{yy})\Bigr],\nonumber
\end{align}
where
\begin{equation}
    S_0 = (C_{11}^0)^2+C_{11}^0C_{12}^0 -2(C_{12}^0)^2,\nonumber
\end{equation}
\begin{equation}
    \tilde{\sigma}_{ij} = N\sum_{m=1}^4\sum_{\Gamma} g_{\Gamma} \braket{O_{\Gamma}^m}\cdot\Bigl(\displaystyle\frac{\partial e_{\Gamma}^m}{\partial e_{ij}}\Bigr).\nonumber
\end{equation}
Figure A1(d) shows the magnetic field induced evolution of electric quadrupolar moments at $m$ = 2. Here $O_\Gamma \equiv \braket{0|O_\Gamma^{m=2}|0}$ and $\ket{0}$ represents the lowest CEF level at each magnetic field. Trigonal crystalline field inherent in the pyrochlore systems induces $O_{20}$ moment even at zero field. The confirmed magnetostriction can be attributable to the magnetically-induced local quadrupolar moments of $O_{20}$, $O_{zx}$ and $O_{22}$.\par
The best fitting parameters used in the present calculation are $C_{11}^0$ = 315 GPa, $C_{12}^0$ = 175 GPa, $g_{22}$ = $-$80 K and $g_{yz}$ = 150 K. The signs of $g_{22}$ and $g_{yz}$ are determined from longitudinal and transverse magnetostrictions. $g_{20}$ cannot be determined since it does not contribute $C_{11}$, $e_{xx}$ and $e_{zz}$.
\bibliography{Tb2Ti2O7}
\end{document}